\newif\ifforarxiv
\forarxivtrue

\ifforarxiv
\documentclass[a4paper,oribibl,11pt]{amsart}

\newtheorem{theorem}{Theorem}[section]
\newtheorem{lemma}[theorem]{Lemma}

\theoremstyle{definition}
\newtheorem{definition}[theorem]{Definition}

\theoremstyle{remark}
\newtheorem{remark}[theorem]{Remark}

\else
\documentclass[a4paper,oribibl,11pt]{llncs}

\fi

\usepackage{fullpage}

\usepackage[latin1]{inputenc}
\usepackage[T1]{fontenc}
\usepackage{lmodern}

\usepackage{multirow}
\usepackage[english]{babel}

\usepackage{amsmath,amssymb,amsfonts}
\usepackage{latexsym}

\usepackage{tikz}

\usepackage[ruled,vlined,algo2e,boxed]{algorithm2e}

\usepackage{enumitem}

\numberwithin{equation}{section}

\newcommand{\F}{\ensuremath{\mathbb{F}}}

\newcommand{\K}{\ensuremath{\mathbb{K}}}
\newcommand{\J}{\ensuremath{\mathcal{J}}}

\newcommand{\OO}{\ensuremath{\mathcal{O}}}

\def\IM{\mathop{\rm{Im}}\nolimits }

\usepackage{fancyhdr}

\usepackage{color}
\usepackage{hyperref}

\hypersetup{ 
  pdfauthor   = {Jean-Gabriel Kammerer, Reynald Lercier, Guénaël Renault}
  pdftitle    = {Computing points on some hyperelliptic curves in deterministic
    polynomial time},
  pdfsubject  = {},
  pdfkeywords = {},
  backref=true, pagebackref=true, hyperindex=true, colorlinks=true,
  breaklinks=true, urlcolor=blue, linkcolor=blue, citecolor=blue, bookmarks=true,
  bookmarksopen=true}

\newenvironment{myproof}[1][\myproofname]{\par
  \normalfont \topsep6pt\relax
  \trivlist
\item[\hskip\labelsep
  \itshape
  #1.]\ignorespaces
}{%
  \endtrivlist\hfill$\square$
}
\providecommand{\myproofname}{Proof}

\begin{document}

\title{Encoding points on hyperelliptic curves over finite fields in
  deterministic polynomial time}

\ifforarxiv

\author{Jean-Gabriel Kammerer}%
\address{ DGA MI, La Roche Marguerite, F-35174 Bruz Cedex, France.  }%
\address{ Institut de recherche math\'ematique de Rennes, Universit\'e de
  Rennes 1, Campus de Beaulieu, F-35042 Rennes Cedex, France.  }%
\email{jean-gabriel.kammerer@m4x.org}

\author{Reynald Lercier}%
\address{ DGA MI, La Roche Marguerite, F-35174 Bruz Cedex, France.  }%
\address{ Institut de recherche math\'ematique de Rennes, Universit\'e de
  Rennes 1, Campus de Beaulieu, F-35042 Rennes Cedex, France.  }%
\email{reynald.lercier@m4x.org}

\author{Gu\'ena\"el Renault}%
\address{LIP6, Universit\'e Pierre et Marie Curie, INRIA/LIP6 SALSA Project-Team, Boite courrier 169, 4 place Jussieu,  F-75252 Paris Cedex 05, France.}%
\email{guenael.renault@lip6.fr}

\fi

\maketitle


\begin{abstract}
  We provide new hash functions into (hyper)elliptic curves over finite
  fields. These functions aims at instantiating in a secure manner
  cryptographic protocols where we need to map strings into points on
  algebraic curves, typically user identities into public keys in
  pairing-based IBE schemes.

  Contrasting with recent Icart's encoding, we start from ``easy to solve by
  radicals'' polynomials in order to obtain models of curves which in turn
  can be deterministically ``algebraically parameterized''.  As a result, we
  obtain a low degree encoding map for Hessian elliptic curves, and for the
  first time, hashing functions for genus 2 curves.  More generally we present
  for any genus (more narrowed) families of hyperelliptic curves with this
  property.

  The image of these encodings is large enough to be ``weak'' encodings in the
  sense of Brier et al., and so they can be easily turned into admissible
  cryptographic encodings.

\end{abstract}

\section{Introduction}
\label{sec:introduction}

Many asymmetric cryptographic mechanisms are based on the difficulty of the
discrete logarithm problem in finite groups. Among these groups, algebraic
curves on finite fields are of high interest because of the small size of keys
needed to achieve good security. Nonetheless it is less easy to encode a
message into an element of the group.

Let $\F_{q}$ be a finite field of odd characteristic $p$, and $H/\F_{q}: y^2 =
f(x)$ where $\deg f = d$ be an elliptic (if $d = 3$ or $4$) or hyperelliptic
(if $d \geqslant 5$) curve, we consider the problem of computing points on $H$
in deterministic polynomial time. In cryptographic applications, computing a
point on a (hyper)elliptic curve is a prerequisite for encoding a message into
its Jacobian group. In this regard, pairing-based cryptosystems do not make
exception.  Boneh-Franklin Identity-Based Encryption scheme~\cite{BonFra2001}
requires for instance to associate to any user identity a point on an elliptic
curve.

In the case of elliptic curves, we may remark that it is enough to compute one
rational point $G$, since we can have other points $t\,G$ from integers $t$ (at
least if $G$ is of large enough order).  To compute such a $G$, one might test
random elements $x \in \F_{q}$ until $f(x)$ is a square.  But without assuming
GRH, we have no guarantee of finding a suitable $x$ after a small enough
number of attempts, and none deterministic algorithm is known for computing
square roots when $p\equiv 1 \bmod 4$. Moreover, encoding $t$ into $t\,G$
voids the security of many cryptographic protocols
\cite{DBLP:conf/crypto/Icart09}.

Maybe a more serious attempt in this direction for odd degrees $d$ is due to
Atkin and Morain~\cite{Atkin:1993:ECP}. They remark that if $x_0$ is any element
of $\F_{q}$ and $\lambda=f(x_0)$, then the point $(\lambda x_0,
\lambda^{(d+1)/2})$ is on the curve $Y^2 = \lambda^{d} f(X/\lambda)$. But again,
the latter can be either isomorphic to the curve or its quadratic twist,
following that $\lambda$ is a quadratic residue or not, and we have no way to
control this in deterministic time.\medskip

In 2006, Shallue and Woestjine~\cite{DBLP:conf/ants/ShallueW06} proposed the
first practical deterministic algorithm to encode points into an elliptic
curve, quickly generalized by Ulas~\cite{ulas07} to the family of
hyperelliptic curves defined by $y^2 = x^n + ax + b$ or $y^2 = x^n + ax^2 +
bx$.  Icart proposed in 2009 another deterministic encoding for elliptic
curves, of complexity $\OO(\log^{2 + o(1)}q)$, provided that the cubic root
function, inverse of $x \mapsto x^3$ on $\F_q^*$, is a group
automorphism. This turns into $q\equiv2 \bmod 3$. This encoding uses
Cardano-Tartaglia's formulae to parameterize the points $(x:y:1)$ on any
elliptic curve $E: x^3 + ax + b = y^2$.

In this paper, we propose a strategy for finding other families with such
properties (Section~\ref{sec:strategy}). As an example, we first show how the
strategy works for genus 1 curves and come to a new encoding map for Hessian
elliptic curves (Section~\ref{sec:an-elliptic-curve}). We then study more
carefully genus 2 curves and exhibit several large families
(Section~\ref{sec:genus-2-case}). Finally for all genus $g \geqslant 2$, we
propose families of hyperelliptic curves which admit an efficient
deterministic encoding function (Section~\ref{sec:determ-param-hyper}),
provided some conditions on $q$ (typically $q=2 \bmod 3$ and $q$ coprime to
$2g+1$).

\begin{remark}
  In the paper, we use indifferently the words ``parameterization'' or
  ``encoding'', even if, strictly speaking, we do not have fully parameterized
  curves. We are aware that these maps are at least improperly
  parameterizations since there might correspond more than one parameter to
  one point. There are numerous points which lie outside the image of our maps
  too.
\end{remark}

\begin{remark}
  Each of our encodings is a \emph{weak encodings} in the sense of
  \cite{conf/crypto/BrierCIMRT10}. Combined with a cryptographic hash
  function, we can thus construct hash functions into the set of rational
  points of these curves that are indifferentiable from a random oracle.
\end{remark}

\section{A strategy}
\label{sec:strategy}
Given a genus $g$, we describe a basic strategy for finding curves of genus $g$
which admit a deterministic encoding for a large subset of their points.

It's worth noting first that only genus $0$ curves are
\emph{rationally} parameterizable. That is, any curve which admits a
rational parameterization shall be a conic, see \cite[Theorem
  4.11]{1564550}.  Encoding maps into higher genus curves shall thus
be \emph{algebraic}. We are then reduced to the parameterization of
roots of polynomials. Hence, the main idea of our general strategy is
to start from polynomials with roots which are easily parameterizable
and then deduce curves with deterministic encoding.

\subsection{Solvable Polynomials}
\label{sec:solvable-polynomials}
Classical Galois theory offers a large family of polynomials with easily
parameterized roots: polynomials with roots that can be written as radicals,
which are polynomials with solvable Galois group. Our strategy is based on
these polynomials.

More precisely, let $f_{\underline{a}}(X)$ be a family of
parameterized polynomials (where $\underline{a}$ denotes a $k$-tuple
$(a_1,a_2,\ldots,a_k)$ of parameters) with solvable Galois group. We
are interested in such parametric polynomials but also in the
parametric radical expression of their roots
$\chi_{\underline{a}}$. For instance $f_{A}(X)=X^2+A$ in degree $2$,
or more interestingly $f_{A,B}(X) = X^3+A\,X+B$ in degree $3$, are
such polynomials with simple radical formulae for their roots. The
former verifies $\chi_A=\sqrt{-A}$ and a root of the second one is
given by the well-known Cardano-Tartaglia's formulae (see
\cite{Cox:GaloisBook}). The application of our general strategy to
this family of degree 3 polynomials with the parameterization of its
roots is described in Section \ref{sec:degree-3-polynomials}.

Let us note that we might use the classical field machinery to
construct new solvable polynomials from smaller ones. Look for
instance at De Moivre's polynomials of degree~$d$: we start from the
degree $2$ field extension $\theta^2+B\theta-A^d$, followed by the
degree $d$ K\"ummer extension $\gamma^d-\theta=0$. Then the element
$X = \gamma-A/\gamma$ is defined in a degree $d$ subfield of the
degree $2d$ extension. The defining polynomial of this extension is
given by the minimal polynomial of $X$, which is equal to the
De Moivre's polynomial,
\begin{displaymath}
X^d + d A X^{d-2} + 2dA^2X^{d-4} + 3dA^3X^{d-6} +\cdots + 2 d A^{(d-1)/2-1} X^3
+dA^{(d-1)/2}X + B\,.
\end{displaymath}
A more straightforward similar construction is to consider K\"ummer
extensions over quadratic (or small degree) extensions, which yields
\begin{math}
  X^{2d} + A X^d + B\,.
\end{math}
From these two specific families of solvable polynomials, we provide, in
Section~\ref{sec:quas-polyn} and~\ref{sec:demoivres-family}, hyperelliptic
curves for all genus $g\geqslant 2$ which admit an efficient deterministic
encoding function.

\subsection{Rational and deterministic parameterizations}

Given a parameterized family of solvable polynomial
$f_{\underline{t}}(X)$, and a genus $g$, we now substitute a rational
fraction $F_i(Y)$ in some variable $Y$ for each parameter $a_i$ in
$\underline{a}$.

Let $\underline{F}(X)$ denote the $k$-tuples of rational fractions
$(F_1(Y),F_2(Y),\ldots, F_k(Y))$. The equation $f_{\underline{F}(Y)}(X)$ now
defines a plane algebraic curve $C$, with variables $(X,Y)$.  The largest are
the degrees in $Y$ of $\underline{F}(Y)$ the largest is (generically) the genus
of $C$. So if we target some fixed genus $g$ for $C$, only few degrees for the
numerators and denominators of $\underline{F}(Y)$ can occur. Since we can
consider coefficients of these rational fractions as parameters
$\underline{a}=(a_1,\ldots,a_{k^{\prime}})$, this yields a family of curves
$C_{\underline{a}}$.

Less easily, it remains then to determine among these
$\underline{F}(Y)$ the ones which yield roots
$\chi_{\underline{F}(Y)}$ which can be computed in deterministic
time. The easiest case is probably when no square root occurs in
the computation of $\chi_{\underline{t}}$, since then any choice for
$\underline{F}(Y)$ will work, at the expense on some constraint on the
finite field. But this is usually not the case, and we might try
instead to link these square roots to some algebraic parameterization of an
auxiliary algebraic curve

\subsection{Minimal Models}

In some case (typically hyperelliptic curves), it is worth to derive from the
equation for $C_{\underline{a}}$ a minimal model (typically of the form
$y^2=g_{\underline{a}}(x)$). In order to still have a deterministic encoding
with the minimal model, we need explicit birational maps
$x=\Lambda_{\underline{a}}(X, Y)$, $y=\Omega_{\underline{a}}(X, Y)$ too.
For hyperelliptic curves, the usual way for this is to work with homomorphic
differentials defined by $C_{\underline{a}}$. This method is implemented in
several computer algebra systems, for instance MAPLE~\cite{maple} or
MAGMA~\cite{Magma:97}.
All in all, we obtain the following encoding for a minimal model $g_{\underline{a}}$:
\begin{itemize}[itemsep=0pt,parsep=0pt,partopsep=0pt,topsep=0pt]
\item Fix some $Y$ as a (non-rational) function of some parameter $t$ so that
  all the square roots are well defined in $\chi_{\underline{F}(Y)}$\,;
\item Compute $X = \chi_{\underline{F}(Y)}$\,;
\item Compute $x=\Lambda_{\underline{a}}(X, Y)$ and $y=\Omega_{\underline{a}}(X, Y)$\,.
\end{itemize}

\subsection{Cryptographic applications}
Once we will have found an encoding, it is important for cryptographic
applications to study the cardinality of the subset of the curve that we
parameterize. This ensures that we obtain convenient weak encodings for
hashing into curves primitives (see~\cite{conf/crypto/BrierCIMRT10}).

We also need to know \emph{in advance} which values of $\F_q$ cannot be encoded
using such functions, in order to deterministically handle such cases. In the
genus 1 as in other sections of our paper, this subset is always quite small
considered to cryptographic sizes (at most several hundred elements) and it
depends only on the once and for all fixed curve parameters, therefore it can be
taken into account and handled appropriately when setting up the
cryptosystem. Furthermore, cryptographic encodings of
\cite{conf/crypto/BrierCIMRT10} make a heavy use of hash functions onto the
finite field before encoding on the curve; the output of the hash function can
then be encoded with overwhelming probability.

In the degree 3 examples given below, as in the higher genus family given in
Section~\ref{sec:demoivres-family}, we always will be able to deduce from the
encoding formulae (sometimes after some resultant computations), a polynomial
relation $P_{\underline{a}}(Y, t)$ between any $Y$ of a point of the image and
its preimages. Then the number of possible preimages is at most the $t$-degree
of $P_{\underline{a}}(Y, t)$. Factorizing $P_{\underline{a}}(Y,t)$ over $\F_q$
gives then precisely the number of preimages.

We detail this process for the genus 1 application of our method in
Section~\ref{sec:cryptographic-use} and sketch how to obtain such a polynomial
in other sections.

\section{Degree 3 polynomials}
\label{sec:degree-3-polynomials}
In this section, we consider degree 3 polynomials. After easy changes
of variables, any cubic can be written in its ``depressed form''
$X^3+3\,A\,X+2\,B$, one root of which is
\begin{displaymath}
  \chi_{A,B} = \sqrt[3]{-B + \sqrt{A^3 + B^2}} - \frac{A}{\sqrt[3]{-B + \sqrt{A^3 + B^2}}}\,.
\end{displaymath}
In order to make use of this root while avoiding square roots, aiming at
(non-rationally) parameterizing curves of positive genus, we first restrict to
finite fields $\F_q$ with $q$ odd and $q\equiv 2 \bmod 3$, so that computing
cubic roots can be done thanks to a deterministic exponentiation to the $e$-th
power, $e=1/3 \bmod q-1$. We then need to consider rational fractions $A$ and $B$
in $Y$ such that the curve ${A(Y)}^{3}+{B(Y)}^{2} - Z^2$ can be parameterized
too.

For non-zero $A$, let $A(Y)=T(Y)$ for some $T$ and $B(Y) = T(Y)S(Y)$ for some
$S$, this problem is then the same as parameterizing the
curve
\begin{equation}\label{eq:2}
  T(Y)+ S^2(Y) = Z^2.   
\end{equation}
This can be done with rational formulae when this curve is of genus 0, or with
non-rational Icart's formulae when this curve is of genus 1. In the case of
irreducible plane curves, this means that $T$ and $S$ are of low
degree. Instead of parameterizing an auxiliary curve, we could have directly
chosen $T$ and $S$ such that $T(Y) + S(Y)^2 = Z(Y)^2$ for some rational
function $Z$. With comparable degrees for $T$ and $S$ as in the rest of the
section, we obtain only genus 0 curves. Thus we have to greatly increase the
degree of $S$ and $T$ in order to get higher genus curves. Those curves then
have high degree but small genus: they have many singularities.\medskip

So, we finally consider in the following degree
3 equations of the form
\begin{equation}
    \label{eq:1}
  {X}^{3}\,+\,3\,T(Y)\,X\,+\,2\,S(Y)T(Y) = 0\,.
\end{equation}
We could have considered the case $A=0$ too, that is polynomials of the form
$f_{B} = {X}^{3}+2B$. Our experiments in genus 1 and genus 2 yield curves that
are isomorphic to hyperelliptic curves of any genus constructed from De Moivre's
polynomials given in Section~\ref{sec:demoivres-family}. We thus do not study
this case further.

\subsection{Genus 1 curves}
\label{sec:an-elliptic-curve}

\subsubsection{Parameterization.}
\label{sec:genus1parametrization}

We made a systematic study of Curves~(\ref{eq:1}) of (generic) genus 1 as a
function of the degree of the numerators and the denominators of the
rational fraction $S(Y)$ and $T(Y)$. Results are in
Tab.~\ref{tab:genus1_degrees}.

\begin{table}[hbtp]
  \centering
  \begin{tabular}{|c|c|cc|c|c|ccc|ccc|c|}\cline{3-13}
    \multicolumn{2}{c|}{} & \multicolumn{11}{c|}{Degrees}\\\hline
    \multirow{2}{*}{$S(Y)$} 
    & Num.   & 2 & 3 & 2 & 0 & 1 & 0 & 1 & 0 & 0 & 0 & 0
    \\
    & Den.   & 0 & 0 & 0 & 1 & 0 & 0 & 0 & 1 & 0 & 1 & 0
    \\\hline
    \multirow{2}{*}{$T(Y)$} 
    & Num.   & 0 & 0 & 1 & 1 & 1 & 2 & 2 & 0 & 0 & 0 & 0
    \\
    & Den.   & 0 & 0 & 0 & 0 & 0 & 0 & 0 & 1 & 2 & 2 & 3
    \\\hline\hline
    \multicolumn{2}{|c|}{Genus of Eq.~(\ref{eq:2})}
             & 1 & 2 & 1 & 1 & 0 & 0 & 0 & 1 & 1 & 1 & 2
    \\\hline
  \end{tabular}\medskip
  \caption{Degrees of $S(Y)$ and $T(Y)$ for
    genus 1 plane curves given by Eq.~(\ref{eq:1})}
  \label{tab:genus1_degrees}
\end{table}

The only case of interest is when $S(Y)$ is a polynomial of degree at most 1
and $T(Y)$ is a polynomial of degree at most 2.  When $q=2\bmod 3$, these
elliptic curves all have a $\F_{q}$-rational $3$-torsion point, coming from
$X=0$.

Elliptic curves with a $\F_{q}$-rational $3$-torsion point are known to have
very fast addition formulae when given in ``generalized'' or ``twisted''
Hessian forms~\cite{FaJo2010,BeKoLa2010}. Since $q=2\bmod 3$, we even restrict
in the following to classical Hessian elliptic curves. \bigskip

Let us start from 
\begin{math}
  S(Y) = 3\,(Y+a)/2,\ T(Y) =-\,Y/3 \,,
\end{math}
that is curves of the type
\begin{equation}
  \label{eq:3}
  C_{0,a}: {Y}^{2}+XY+aY={X}^{3}\,,\ a\neq 0, 1/27\,.
\end{equation}
Then, the conic
\begin{math}
  S^2(Y)+T(Y) = 9/4\,{Y}^{2}+ \left( 9/2\,a-1/3 \right) Y+9/4\,{a}^{2} = {Z}^{2}\,
\end{math}
can be classically parameterized ``by line'' as
\begin{displaymath}
Y={\frac {12\,{t}^{2}-27\,{a}^{2}}{36\,t-4+54\,a}}\,,\ 
Z={\frac {36\,{t}^{2}+ \left( -8+108\,a \right) t+81\,{a}^{2}}{72\,t-8+108\,a}}\,,
\end{displaymath}
so that $X= \Delta/6+2Y/\Delta$ where $\Delta= \sqrt [3]{36 Y \left(
    3\,Y+3\,a+2\,Z \right) }$.

Besides, Curve~(\ref{eq:3}) is birationally equivalent to the
Hessian model
\begin{equation}\label{eq:4}
  E_{d}: {x}^{3}+{y}^{3}+1 = 3\,dxy,\ d \neq 1,
\end{equation}
with $a=({ {{d}^{2}+d+1})/{ 3\left( d+2 \right) ^{3}}}$
and
\begin{equation}\label{eq:10}
x = {\frac { 3\,\left( d+2 \right) ^{2} \left( Y\,(d+2)+X\right) }{3\,\left(
      d+2 \right) ^{2}X+{d}^{2}+d+1}}\,,\  
y = - {\frac {{d}^{2}+d+1+3\, \left( d+1 \right)  \left( d+2 \right) ^{2}X+
3\, \left( d+2 \right) ^{3}Y}{3\, \left( d+2 \right) ^{2}X+{d}^{2}+d+1
}}\,.
\end{equation}
\smallskip

The only remaining case is $d=-2$, that is the Hessian curve $E_{-2}$ (the
quadratic twist of the curve $E_0$, both have their $j$-invariant equal to
0). This curve is for instance isomorphic to a curve of the type~\eqref{eq:1}
with $S = (1-7\,Y)/4$ and $T = -26\,(3\,{Y}^{2}+1)/27$. We might use this to
parameterize $E_{-2}$, but it is much simplier to start from the curve
$Y^2+Y=X^3$, which can be much more easily parameterized with $Y=t$, $X =
\sqrt [3]{t^2+t}$. This curve is isomorphic to $E_{-2}$ with $x={(
  {X+1})/({X+Y})}$, $y={ ({-Y+X-1})/({X+Y})}$.

We summarize these calculations in Algorithm~\ref{algoGenus1}.
\begin{figure}[htbp]
  \centering
  \parbox{0.8\textwidth}{%
    \begin{footnotesize}%
      \begin{algorithm2e}[H]%
        \SetKwInOut{Input}{input} \SetKwInOut{Output}{output} %
        \Input{A Hessian elliptic curve $E_{d}/\F_q: {x}^{3}+{y}^{3}+1 =
          3\,dxy,\ d \neq 1$, and $t \in \F_q$.}%
        \Output{A point $(x_t:y_t:1)$ on $E_{d}$.}%
        \BlankLine%
        \If(\tcc*[f]{$t \ne 0$}){ $d = -2$}{ %
          {$Y := t$; $X := (t+t^2)^{1/3 \bmod q-1}$}\;%
          {$x_t := (X+1)/(X+Y)$; $y_t := (-Y+X-1)/(X+Y)$}\;%
          \KwRet{$(x_t:y_t:1)$}%
        }%
        $a := \displaystyle \frac {{d}^{2}+d+1}{3\, \left( d+2 \right) ^{3}}$%
        {\tcc*{$t \ne {\displaystyle \frac { ( 2\,d+1 ) ( {d}^{2}+d+7 ) }{ 18\, \left( d+2
                \right) ^3}}$}}%
        \eIf{$ t = \pm 3a/2$} { %
          $Y := 0$; $X := 0$\;%
        }%
        (\tcc*[f]{$Y \not = 0$}) {%
          $Y := {\displaystyle \frac {12\,{t}^{2}-27\,{a}^{2}}{36\,t+54\,a-4}}$;\ %
          $\Delta := \left( 36\,Y \left(2\,t+3\,a \right) \right)^{1/3
              \bmod q-1}$;\ %
            $X := \Delta/6+2\,{{Y}/{\Delta}}$\;%
        }%
        {$x_t := {\displaystyle \frac {3\, \left( d+2 \right) ^{2} \left( Y \left( d+2
                \right) +X \right) }{3\, \left( d+2 \right)
              ^{2}X+{d}^{2}+d+1}}$~;\ %
          $y_t := -{\displaystyle \frac {3\, \left( d+1 \right) \left( d+2 \right) ^{2}X+3\,
              \left( d+2 \right) ^{3}Y+{d}^{2}+d+1}{3\, \left( d+2 \right)
              ^{2}X+{d}^{2}+d+ 1}}$}\;%
        \KwRet{$(x_t:y_t:1)$}%
        \label{algoGenus1} \caption{HessianEncode}
      \end{algorithm2e}
    \end{footnotesize}
  }
  \caption{Encoding on Hessian elliptic curves}
  \label{fig:algoGenus1}
\end{figure}
\medskip

In addition, we have proved what follows.
\begin{theorem}
  Let $\F_q$ be the finite field with $q$ elements. Suppose $q$ odd and $q
  \equiv 2 \mod 3$. 
  Let $E_{d}/\F_q~$ be the elliptic curve defined by Eq.~(\ref{eq:4}).

  Then Algorithm~\ref{algoGenus1} computes a deterministic encoding
  $e_{d}$ to $E_{d}$, from $\F_q^*$ if $d=-2$ and from $\F_q\setminus
  \left\{{\frac { \left( 2\,d+1 \right) \left( {d}^{2}+d+7 \right) }{
        18\, \left( d+2 \right) ^{3}}} \right\}$ otherwise, in time
  $\OO(\log^{2+o(1)} q)$.
\end{theorem}

A way of quantify the number of curves defined by Eq.~(\ref{eq:4}) is
to compute their $j$-invariant. Here, we obtain
\begin{equation}
  \label{eq:5}
  j_{E_{d}} = 27\,d^3\, {\frac {\left( d+2 \right) ^{3} \left( {d}^{2}-2\,d+4
 \right) ^{3}}{ \left( d-1 \right) ^{3} \left( {d}^{2}+d+1 \right) ^{3
}}}\,.
\end{equation}
When $q \equiv 2 \mod 3$, there are exactly $\lfloor q/2 \rfloor$ distinct
such invariants. Additionally, one can show that there exists $q-1$ distinct
$\F_q$-isomorphic classes of Hessian elliptic curves
(see~\cite{FaJo2010}).

\subsubsection{Cardinality of the image.}
\label{sec:cryptographic-use}

It is obvious to see that $|\IM e_{-2}|=q-1$, simply because $Y=t\neq 0$. Now,
determining $|\IM e_{d}|$ for $d\ne 1, -2$ needs some more work, but can still
be evaluated exactly.

\begin{theorem}
  Let $d\ne 1, -2$, then $|\IM e_{d}| = (q+1)/2$ if $(d-1)/(d+2)$ is a
  quadratic residue in $\F_q$ and $|\IM e_{d}| = (q-1)/2$ otherwise.
\end{theorem}

\begin{myproof}
  Let $(x:y:1)$ be a point on $E_{d}$, then there exists a unique point
  $(X:Y:1)$ on $C_{0,a}$ sent by Isomorphism~\eqref{eq:10} to $(x:y:1)$.

  Viewed as a polynomial in $t$, the equation
  $12\,{t}^{2}-36\,Yt-54\,Ya-27\,{a}^{2}+4\,Y$ has 0 or 2 solutions except
  when $27\,{Y}^{2}+ \left( -4+54\,a \right) Y+27\,{a}^{2}=0$.  The latter has
  no root if $1-27\,a = { { \left( d-1 \right) ^{3}}/{ \left( d+2 \right)
      ^{3}}}$ is a quadratic non-residue, and two distinct roots denoted $Y_0$
  and $Y_1$ otherwise (if $a = 1/27$, the curve $C_{0,a}$ degenerates into a
  genus 0 curve).\medskip

  Let us summarize when $\left( d-1 \right)/ \left( d+2 \right)$ is a
    quadratic residue in $\F_q$.
  \begin{itemize}[itemsep=0pt,parsep=0pt,partopsep=0pt,topsep=0pt]
  \item ($1$ element) If $t \in \left\{{\frac { \left( 2\,d+1 \right) \left(
            {d}^{2}+d+7 \right) }{ 18\, \left( d+2 \right) ^{3}}}\right\}$, then
    $t$ is not encodable by $e_{d}$;
  \item ($2$ elements) If $t \in \{ \pm\,{\frac
        {{d}^{2}+d+1}{ 2\, \left( d+2 \right) ^{3}}} \}$, then $e_d(t) = (0: -1: 1)$;
  \item ($2$ elements) If $t_i$ is a (double) root of $12\,{t}^{2}- \left(
      36\,t-4+54\,a \right) Y_i-27\,{a}^{2}$ with $i = 0, 1$, we obtain two
    distinct points $e_{d}(t_i) = (x_{t_i}:y_{t_i}:1)$;
  \item ($q-5$ elements) Else, for each remaining $t$, there exists
    exactly one other $t'$ such that $e_{d}(t) = e_{d}(t') = (x_t:y_t:1)$.
  \end{itemize}
  We thus obtain $(q-5)/2 + 2 +2 = (q+1)/2$ distinct rational points on the
  curve.  Similarly if $\left( d-1 \right)/\left( d+2 \right)$ is a
    quadratic non-residue in $\F_q$, we obtain $(q-1)/2$ distinct rational
    points on $E_{d}$.
\end{myproof}
  
\subsubsection{Related work.}
\label{sec:comparison-icart}
Compared to Icart's formulae~\cite{DBLP:conf/crypto/Icart09}, this
encoding has two drawbacks of limited practical impact:
\begin{itemize}[itemsep=0pt,parsep=0pt,partopsep=0pt,topsep=0pt]
\item it does not work for any elliptic curves, but only for Hessian curves;
\item the subset of the curve which can be parameterized is slightly
  smaller than in Icart's case: we get $\simeq q/2$ points
  against approximately $5/8 \#E  \pm \lambda \sqrt{q}$. 
\end{itemize}

{\noindent}Nonetheless, it has three major practical advantages: 
\begin{itemize}[itemsep=0pt,parsep=0pt,partopsep=0pt,topsep=0pt]
\item recovering the parameter $t$ from a given point $(x:y:1)$ is much
  easier: we only have to find the roots of a degree 2 equation
  instead of a degree 4 one;
\item the parameter $t$ only depends on $y$: we can save half of the bandwidth
  of a protocol by sending only $y$ and not the whole point $(x:y:1)$;
\item $y_t$ is computable using only simple (rational) finite field operations:
  no exponentiation is required, but it carries the whole information on the
  encoded point\footnote{For example, we could imagine that a limited power device
  computes the encoded $y$ and sends it to an other device specialized in curve
  operations, which in turn computes the associated $x$ and realizes the group
  operations.}.
\end{itemize}

\subsection{Genus 2 curves}
\label{sec:genus-2-case}

\subsubsection{Parameterizations.}
\label{sec:genus2parametrization}

In the same spirit as in Section~\ref{sec:an-elliptic-curve}, we made
a systematic study of Curves~(\ref{eq:1}) of (generic) genus 2 as a function of
the degree of the numerators and the denominators of the rational
fraction $S(Y)$ and $T(Y)$. Results are in
Tab.~\ref{tab:genus2_degrees}. 

\begin{table}[hbtp]
  \centering
  \begin{tabular}{|c|c|cccc|cc|ccccc|cc|c|ccc|}\cline{3-19}
    \multicolumn{2}{c|}{} & \multicolumn{17}{c|}{Degrees}\\\hline
    \multirow{2}{*}{$S(Y)$} 
    & Num.   & 2 & 0 & 1 & 2 & 2 & 2 & 1 & 1 & 0 & 1 & 1 & 1 & 1 & 2 & 0 & 0 & 0
    \\
    & Den.   & 1 & 2 & 2 & 2 & 0 & 1 & 1 & 0 & 1 & 1 & 1 & 0 & 0 & 0 & 0 & 0 & 0
    \\\hline
    \multirow{2}{*}{$T(Y)$} 
    & Num.   & 0 & 0 & 0 & 0 & 0 & 0 & 0 & 1 & 1 & 1 & 1 & 0 & 1 & 0 & 1 & 2 & 2
    \\
    & Den.   & 0 & 0 & 0 & 0 & 1 & 1 & 1 & 1 & 1 & 0 & 1 & 2 & 2 & 2 & 2 & 1 & 2
    \\\hline\hline
    \multicolumn{2}{|c|}{Genus of Eq.~(\ref{eq:2})}
             & 1 & 1 & 1 & 1 & 2 & 2 & 1 & 1 & 1 & 1 & 1 & 2 & 2 & 3 & 1 & 1 & 1
    \\\hline
  \end{tabular}\medskip
  \caption{Degrees of $S(Y)$ and $T(Y)$ for
    genus 2 plane curves given by Eq.~(\ref{eq:1})}
  \label{tab:genus2_degrees}
\end{table}

We can see that there are three cases of interest:
\begin{itemize}[itemsep=0pt,parsep=0pt,partopsep=0pt,topsep=0pt]
\item $S(Y)$ and $T(Y)$ be both a rational fraction of degree 1~;
\item $S(Y)$ be a rational fraction of degree 2 and $T(Y)$ be a constant~;
\item $S(Y)$ be a constant and $T(Y)$ be a rational fraction of degree 2.
\end{itemize}
We now study the two first cases. We omit the third one because it
turns out that it yields curves already obtained in the second case.

\subsubsection{$S(Y)$ and $T(Y)$  rational fractions of degree 1.}
\label{sec:sy-ty-be}

Let $S(Y) = {({\alpha Y+\beta })/({\gamma Y+\delta })}$ and $T(Y) =
{({\varepsilon Y+\varphi })/({\mu Y+\nu })}$, then Curve~(\ref{eq:1}) is
birationally equivalent to curves of the form $y^2 / {d}^{2} =
( {x}^{3}+3\,ax+2\,c ) ^{2}+8\,b{x}^{3}$ where
 \begin{displaymath}
   a={\frac {\delta \varepsilon -\gamma \varphi }{\delta \mu -\gamma \nu
     }},\ %
   b={\frac { ( \alpha \delta -\gamma \beta  )  ( \mu
         \varphi -\varepsilon \nu  ) }{ ( \delta \mu -\gamma \nu
       ) ^{2}}}\,, %
   c={\frac {\beta \varepsilon -\alpha \varphi }{\delta \mu
       -\gamma \nu }}\ %
   \text{ and }\ d= ( \delta \mu -\gamma \nu  )\,.
 \end{displaymath}
 Many of theses curves are isomorphic to each other and, without any
 loss of generality, we can set $c=1$ and $d=1$. We thus finally
 restrict to
\begin{math}
  S(Y) = -Y,\ T(Y) = {({{a}^{2}Y}+a ) / ({aY+b+1})}\,,
\end{math}
so that, when $4\,{a}^{6}{b}^{3}-{b}^{3} \left( {b}^{2}+20\,b-8
\right) {a}^{3}+4\,{b}^{3} \left( b+1 \right) ^{3}\neq 0$,
Curve~(\ref{eq:1}) is birationally equivalent to the Weierstrass model
of a genus 2 curve,
\begin{equation}\label{eq:7}
  H_{1,a,b}: {y}^{2} = ( {x}^{3}+3\,ax+2 ) ^{2}+8\,b{x}^{3}\,,
\end{equation}
with $x = X$ and $y = -4\,aY+{X}^{3}+3\,aX-2$.\medskip

Besides, Curve
\begin{equation}\label{eq:6}
  S^2(Y)+T(Y) = {Y}^{2}+{ ( a^2\,Y+a)/({aY+1+b})} = Z^2
\end{equation}
is birationally equivalent to the Weierstrass elliptic curve
\begin{multline}\label{eq:11}
  {V}^{2}={U}^{3}%
  +( -{a}^{6}+2\, ( b+1 ) ( 2 \,b-1 )
    {a}^{3}- ( b+1 ) ^{4} )\,\frac{U}{3}\\%
  +\frac{1}{27}\,(2\,{a}^{9}+3\, ( 2-2\,b+5\,{b}^{2} ) {a}^{6}-6\, (
    2\, b-1 ) ( b+1 ) ^{3}{a}^{3}+2\, ( b+1
  ) ^{6 })\,.
\end{multline}
The latter can now be parameterized with Icart's method. This yields
\begin{displaymath}
  U = \frac{1}{6}\,\sqrt [3]{{\frac
      {{2\delta}}{{t}^{2}}}}+\frac{{t}^{2}}{3},\ %
  V =\frac{1}{6}\,\sqrt [3]{{{{2\delta}}{{t}}}}%
  + \frac{{t}^{3}}{6}+\frac{1}{6t}\, (-{a}^{6}+2\, ( b+1 )  ( 2\,b-1
  ) {a}^{3}- ( b+1 ) ^{4})
\end{displaymath}
with
\begin{multline*}
  \delta = -{t}^{8}+ ( -12\, ( b+1 ) ( 2\,b-1 ) {a}^{
    3}+6\,{a}^{6}+6\, ( b+1 ) ^{4} ) {t}^{4} + ( 12 \, (
  2\,b-5\,{b}^{2}-2 ) {a}^{6}-8\, ( b+1 ) ^{6}\\%
  -8\,{a}^{9}+24\, ( 2\,b-1 ) ( b+1 ) ^{3}{a}^{3 } ) {t}^{2}+3\, (
  {a}^{6}-2\, ( b+1 ) ( 2 \,b-1 ) {a}^{3}+ ( b+1 ) ^{4} ) ^{2}
\end{multline*}
Now, back by the birational change of variables between Curve~(\ref{eq:11})
and Curve~(\ref{eq:6}), we get $Y$ and $Z$ from $U$ and $V$ (\textit{cf.}
Algorithm~\ref{algo:genus2-sy-ty-rational} for precise formulae).  Let
now
\begin{math}
  \Delta = \sqrt [3]{T(Y) (  {Z}-S(Y) ) }\,,%
\end{math}
then $X={ {\Delta}-T(Y)/{\Delta}}$.

\begin{figure}[htbp]
  \centering
  {\parbox{0.9\textwidth}{%
      \begin{footnotesize}%
        \begin{algorithm2e}[H]%
          \dontprintsemicolon
          \SetKwInOut{Input}{input} \SetKwInOut{Output}{output} %
          \Input{A curve $H_{1, a,b}$ defined by Eq.~(\ref{eq:7})
            on $\F_q$, an element $t\in\F_q\setminus {\mathcal S}_1 $}%
        \Output{A point $(x_t:y_t:1)$ on $H_{1, a, b}$}%
        \BlankLine%
        $\!\begin{array}[b]{rcl}%
          \delta &:=& %
          -{t}^{8}+ ( -12\, ( b+1 ) ( 2\,b-1 ) {a}^{ 3}+6\,{a}^{6}+6\, (
          b+1 ) ^{4} ) {t}^{4} + ( 12 \, ( 2\,b-5\,{b}^{2}-2 )
          {a}^{6}-8\, ( b+1 ) ^{6}\\&& %
          -8\,{a}^{9}+24\, ( 2\,b-1 ) ( b+1 ) ^{3}{a}^{3 } ) {t}^{2}+3\,
          ( {a}^{6}-2\, ( b+1 ) ( 2 \,b-1 ) {a}^{3}+ ( b+1 ) ^{4} ) ^{2};%
        \end{array}$\;%
        $U~:= (({{
            {{2\delta}}/{{t}^{2}}}})^{1/3 \bmod q-1}+{{2t}^{2}})/{6}$;\;%
        $V~:=({{{{2\delta}}{{t}}}})^{1/3 \bmod q-1}/6%
        + {{t}^{3}}/{6} + (-{a}^{6}+2\, ( b+1 )  ( 2\,b-1
        ) {a}^{3}- ( b+1 ) ^{4}) / 6t$;\;%
        $W~:= -3\,Ua+a (  ( b+1 ) ^{2}+{a}^{3} )$;\ %
        $Y~:= (3\, ( b+1 ) U+ ( 2\,b-1 ) {a}^{3}- ( b+1) ^{3}) / W$;\ %
        $Z~:= 3\,V/W$;\;%
        $T ~:= {({{a}^{2}Y}+a ) / ({aY+b+1})}$;\ %
        $\Delta~:= ( T ( Z+Y )  ) ^{1/3 \bmod q-1}$;\;%
        $x_t~:={ {\Delta}-T/{\Delta}}$;\ %
        $y_t~:= -4\,aY+{X}^{3}+3\,aX-2$;\;%
        \KwRet{$(x_t:y_t:1)$}
        \label{algo:genus2-sy-ty-rational} \caption{Genus2Type1Encode}
      \end{algorithm2e}
    \end{footnotesize}
  }}
\caption{Encoding on genus 2 curves (of the type 1)}
\label{fig:algo:genus2-sy-ty-rational}
\end{figure}
\medskip
  
So, we obtain the following theorem.
\begin{theorem}\label{th:genus2_II}
  Let $\F_q$ be the finite field with $q$ elements. Suppose $q$ odd and $q
  \equiv 2 \mod 3$. 
  Let $H_{1,a,b}/\F_q~$ be the hyperelliptic curve of genus 2 defined
  by Eq.~(\ref{eq:7}).

  Then, Algorithm~\ref{algo:genus2-sy-ty-rational} computes a
  deterministic encoding $e_{1, a, b}: \F_q^*\setminus S \rightarrow
  H_{1, a, b }$, where ${\mathcal S}_1$ is a subset of $\F_q$ of size
  at most $74$, in time $\OO(\log^{2+o(1)} q)$.
\end{theorem}

\begin{myproof}
The previous formulae define a deterministic encoding provided that
$t$, $W$, ${aY+b+1}$ and $\Delta$ are not 0. 

The condition $W=0$ yields a polynomial in $t$ of degree $8$, we thus
have at most 8 values for which $W=0$. Similarly, the condition
${aY+b+1}=0$ yields at most 8 additional values for which $W=0$.

Now $\Delta = 0$ if and only if $T=0$ or $Z=-Y$. The condition $T=0$
yields 8 additional values. Similarly, the condition $Z+Y=0$ yields a
polynomial in $t$ of degree $10$, we thus have in this case at most 18
values for which $\Delta=0$.

The total number of field elements which cannot be encoded finally
amounts to at most $35$.
\end{myproof}

\subsubsection{Cardinality of the image.}
\label{sec:card_genus2_1}
Let $(X, Y)$ be a rational point on a $C_{1, a, b, c}$ curve, let $t$
be a possible preimage of $(X, Y)$ by our encoding $e_{1, a,b}$. Then
there exists a polynomial relation in $Y$ and $t$ of degree at most
8 in $t$ (\textit{cf.}  Algorithm~\ref{algo:genus2-sy-ty-rational}).
 Hence $(X, Y)$ has at most $8$ preimages by
$e_{1, a, b}$. Therefore, $|\IM e_{1, a, b}| \geqslant ({q-35})/{8}$.

\subsubsection{Number of curves.}
Igusa invariants of these curves are equal to
\begin{small}
  \begin{displaymath}
    \begin{array}{rcl}
      J_2 & = & 2^6\,3\,(-9\,{a}^{3}+4\,{b}^{2}+4\,b-9)\,,\\
      J_4 & = & 2^{10}\,3\, (-9\,b ( 4\,b-15 ) {a}^{3}+4\,b ( b+1 )  (
      2 \,{b}^{2}+2\,b-27 ))\,,\\
      J_6 & = & 2^{14}\, (729\,{a}^{6}{b}^{2}-216\,{b}^{2} \left(
        2\,{b}^{2}+3\,b+21 \right) {a}^{3}+16\,{b}^{2} \left(
        4\,{b}^{2}+4\,b+81 \right)  \left( b+1 \right) ^{2})\,,\\
      J_8 & = & 2^{18}\,3\, 
      (-6561\,{a}^{9}{b}^{2}+2916\,{b}^{2} \left( -7+{b}^{2}+13\,b \right) {a
      }^{6}\\&&%
      -144\,{b}^{2} \left( 4\,{b}^{4}+63\,{b}^{3}+450\,{b}^{2}-149\,b-
        810 \right) {a}^{3}\\&&%
      +64\,{b}^{2} \left( {b}^{4}+2\,{b}^{3}+154\,{b}^{2}
        +153\,b-729 \right)  \left( b+1 \right) ^{2})\,,\\
      J_{10} & = & 2^{28}\,3^6\, (4\,{a}^{6}{b}^{3}-{b}^{3} \left(
        {b}^{2}+20\,b-8 \right) {a}^{3}+4\,{b}^{3} \left( b+1 \right)
      ^{3})\,.
  \end{array}
  \end{displaymath}
\end{small}
The geometric locus of these invariants is a surface of dimension 2
given by a homogeneous equation of degree 90 (which is far too large
to be written here).  Consequently, Eq.~(\ref{eq:7}) defines
$\OO(q^2)$ distinct curves over $\F_q$.

\subsubsection{$S(Y)$ be a rational fraction of degree 2.}
\label{sec:sy-be-rational}
Let now $S(Y)=({\alpha {Y}^{2}+\beta Y+\gamma })\,$ $/\,({\delta
    {Y}^{2}+\varepsilon Y+\varphi})$ and $T(Y)=\kappa$,
then Curve~(\ref{eq:1}) is birationally equivalent to curves of the
form $y^2/\lambda  = ( {x}^{3}+3\,\mu x+2\,a ) ^{2}+4\,b$ where
\begin{displaymath}
  \lambda  = {\varepsilon}^{2}-4\,\varphi\delta \,,\ %
  \mu  = \kappa\,,\ %
  a = {{\frac{\kappa}{\lambda } ( \varepsilon\beta -2\,\delta \gamma -2\,\varphi\alpha  ) }}\ %
  \text{ and }b = {\frac{{\kappa}^{2}}{\lambda } ({\beta }^{2} -4\,\alpha \gamma  )}-{a}^{2}\,.
\end{displaymath}
Many of theses curves are isomorphic to each other and, without any
loss of generality, we can set $\lambda$ and $\mu$ to be either any
quadratic residues (for instance $\lambda, \mu=1$) or any
non-quadratic residues (for instance $\lambda, \mu=-3$ because $q=2
\bmod 3$). \medskip

We finally arrive to
\begin{displaymath}
  S(Y) = {\frac { \lambda ( a-u ) {Y}^{2}-4\,vY-4\,(a+\,u)}{\mu (
      \lambda{Y}^{2}-4 ) }}\ \text{and }\ T(Y) = \mu\,,
\end{displaymath}
where
\begin{math}
  u= {{{\mu}^{3}}/{2\,w}} -w/2 - a
\end{math}
for some $w\in\F_q^*$.  Then, when ${b}^{3}{\lambda}^{10} ( {\mu
}^{6}+2\,{\mu }^{3}{a}^{2}-2\,b{\mu }^{3}+{a}^{4}+2 \,b{a}^{2}+{b}^{2}
)\neq 0$, Curve~(\ref{eq:1}) is birationally equivalent to the
Weierstrass model of a genus 2 curve,
\begin{equation}\label{eq:9}
  H_{2,\lambda,\mu,a, v,w}: {y}^{2} / \lambda= ( {x}^{3}+3\,\mu x+2\,a )
  ^{2}+4\,b\,,
\end{equation}
where $b= {v}^{2}/\lambda-{u}^{2}$ for some $v$ in $\F_q$, $x = X$ and
$y = \lambda\, ({X}^{3}/2+3\,\mu X/2+a-u)\, Y-2\,v$.

We may remark that computing $v$ and $w$ from $b$ is the same as
computing a point $(v:w:1)$ on the elliptic curve ${v}^{2}/\lambda -
({{{\mu}^{3}}/{2\,w}} -w/2 - a)^2 -b=0$. This can be done in
deterministic time from Icart's formulae when one can exhibit a
$\F_q$-rational bilinear change of variable between this curve and a
cubic Weierstrass model, typically when $\lambda=1$ (but no more when
$\lambda=-3$).\medskip

Besides, let $z=w/2+{ {{r}^{3}}/{2w}}$ and thus
$(u+a)^{2}+{r}^{3}={z}^{2}$, then
\begin{multline}
\label{eq:8}
  {\mu^2 ( \lambda{Y}^{2}-4 )^2 }(S(Y)^2+T(Y)) = -{\lambda}^{2} (
    4\,ua-{z}^{2} ) {Y}^{4}-8\,\lambda v ( a-u ) {Y}^{
3}\\
-8\,\lambda ( 4\,{\mu}^{3}-3\,{z}^{2}-2\,b+6\,ua+4\,{a}^{2} ) {Y}^{2}+
32\,v ( u+a ) Y+16\,{z}^{2} = Z^2
\end{multline}
is birationally equivalent to the Weierstrass elliptic curve
\begin{multline}\label{eq:12}
  {V}^{2} = {U}^{3}+ 2^8 {\lambda }^{2}( -\,{\mu }^{6}+\, ( b-2\,{a}^{2} ) {\mu }^{3}-\,
  ( {a}^{2}+b ) ^{2} ) U/3+\\%
  2^{12} {\lambda }^{3}(2\,{\mu }^{9}+ \, ( 6\,{a}^{2}-3\,b ) {\mu }^{6}-3\, ( {a}^{2}+b )
  ( b-2\,{a}^{2} ) {\mu }^ {3}+2\, ( {a}^{2}+b ) ^{3}) /3^3\,.
\end{multline}
The latter can now be parameterized with Icart's method. This yields
\begin{displaymath}
  U = \frac{1}{6}\,\sqrt [3]{{\frac {{2\delta}}{{t}^{2}}}}+\frac{{t}^{2}}{3},\
  \ V =\frac{1}{6}\,\sqrt [3]{{{{2\delta}}{{t}}}}
  + \frac{{t}^{3}}{6}+
 128\, ( -{\mu }^{6}+ ( b-2\,{a}^{2} ) {\mu }^{3}- ( b+{a}^{2
} ) ^{2} ) \frac{{\lambda }^{2}}{3t}
\end{displaymath}
with
\begin{multline}
  \delta = -{t}^{8}+ 2^9\,3\,( \,{\mu }^{6}+\, ( -b+2\,{a}^{2} ) {\mu }
  ^{3}+\, ( {a}^{2}+b ) ^{2} ) {\lambda }^{2}{t}^{4}+\\%
  2^{14}( -2\,{\mu }^{9}- ( 6\,{a}^{2}-3\,b ) {\mu }^{6} +3\, ( {a}^{2}+b )
  ( b-2\,{a}^{2} ) {\mu }^{3} -2\, ( {a}^{2}+b ) ^{3} )
  {\lambda }^{3}{t}^{2}+\\%
  2^{16}\,3\,(
  \,{\mu }^{12}+\, ( -2\,b+4\,{a}^{2} ) {\mu }^{9}+ \, (
  3\,{b}^{2}+6\,{a}^{4} ) {\mu }^{6}+2\, ( {a}^{2}+b ) ^{2} (
  -b+2\,{a}^{2} ) {\mu }^{3}+\, ( {a}^{2}+b ) ^{4} ) {\lambda }^{4}\,.
\end{multline}
Again, back by a birational change of variables between
Curves~\eqref{eq:12} and~\eqref{eq:8}, we get $Y$ and $Z$ from
$U$ and $V$ (\textit{cf.} Algorithm~\ref{algo:genus2-sy-rational} for
precise formulae).  Let now
\begin{math}
  \Delta = \sqrt [3]{T(Y) \left( { {Z}/{\mu ( \lambda{Y}^{2}-4 ) }}-S(Y)
    \right) }\,,%
\end{math}
then $X={ {\Delta}-T(Y)/{\Delta}}$\,.

\begin{figure}[htbp]
  \centering
  \hspace*{-0.5cm}{\parbox{1.05\textwidth}{%
      \begin{footnotesize}%
        \begin{algorithm2e}[H]%
          \dontprintsemicolon
          \SetKwInOut{Input}{input} \SetKwInOut{Output}{output} %
          \Input{A curve $H_{2,\lambda,\mu,a, v,w}$ defined by
            Eq.~\eqref{eq:9} on $\F_q$, an element $t\in\F_q\setminus {\mathcal S}_2$.}%
          \Output{A point $(x_t: y_t:1)$ on $H_{2,\lambda,\mu,a, v,w}$}%
          \BlankLine%
          $u~:=-{ ({2\,aw+{w}^{2}-{r}^{3}})/{2w}}$; $b~:={
            {{v}^{2}}/{l}}-{u}^{2}$; $z~:=({ {{w}^{2}+{r}^{3}})/{2w}}$;\;
          $\!\begin{array}[b]{rcl}%
            \delta &:=& -{t}^{8}+ 2^9\,3\,( \,{\mu }^{6}+\, ( -b+2\,{a}^{2} )
            {\mu } ^{3}+\, ( {a}^{2}+b ) ^{2} ) {\lambda }^{2}{t}^{4}+\\%
            &&2^{14}( -2\,{\mu }^{9}- ( 6\,{a}^{2}-3\,b ) {\mu }^{6} +3\, (
            {a}^{2}+b ) ( b-2\,{a}^{2} ) {\mu }^{3} -2\, ( {a}^{2}+b ) ^{3} )
            {\lambda }^{3}{t}^{2}+\\%
            &&2^{16}\,3\,( \,{\mu }^{12}+\, ( -2\,b+4\,{a}^{2} ) {\mu }^{9}+
            \, ( 3\,{b}^{2}+6\,{a}^{4} ) {\mu }^{6}+2\, ( {a}^{2}+b ) ^{2} (
            -b+2\,{a}^{2} ) {\mu }^{3}+\, ( {a}^{2}+b ) ^{4} ) {\lambda
            }^{4}; \end{array}$\;%
          $U~:= (({2\,\delta}/{t^2})^{1/3 \bmod q-1}+2{t}^{2})/6$;\;%
          $V~:= (2\delta t)^{1/3 \bmod q-1}/6 + {t}^{3}/6+ 128\, ( -{\mu
          }^{6}+ ( b-2\,{a}^{2} ) {\mu }^{3}- ( b+{a}^{2 } ) ^{2} ) {{\lambda
            }^{2}}/{3t}$;\;%
          $\!\begin{array}[b]{rcl}%
            W &:=&-9\,{U}^{2}-48\,\lambda (
            -3\,{z}^{2}-2\,b+6\,ua+4\,{a}^{2}+4\,{\mu }^{3} ) U+256\, ( -4\,
            {\mu }^{6}+ ( 6\,{z}^{2}+{a}^{2}-12\,ua+4\,b ) {\mu }^{3}+\\&&%
            ( b+{a}^{2} ) ( 5\,{a}^{2}+6\,ua-b-3\,{z}^{2} ) ) {\lambda
            }^{2}; \end{array}$\;%
          $\!\begin{array}[b]{rcl}%
            Y &:=& (-288\,v ( u+a ) U-72\,zV+1536\,\lambda v (
            bu+{a}^{3}-2\,{\mu}^{3}u+ab+a{\mu}^{3}+u{a}^{2} ))/W; \end{array} $\;%
          $\!\begin{array}[b]{rcl}%
            Z &:=& -(-324\,z{U}^{4}+ ( 6912\,\lambda {\mu}^{3}z+1728\,\lambda z
            ( -3\,{z}^{2}-2\,b+6\,ua+4\,{a}^{2} ) ) {U}^{3}-2592\,v ( u+a )
            {U}^{2}V\\&&%
            +( -27648\,{\lambda }^{2}z ( b+{a}^{2} ) (
            2\,{a}^{2}+6\,ua-4\,b-3\,{z}^{2} ) +193536\,{\lambda
            }^{2}z{\mu}^{6}-27648\,{\lambda }^{2} z (
            -5\,{a}^{2}-12\,ua+6\,{z}^{2}+7\,b ) {\mu}^{3} ) {U}^{2}\\&&%
            +( 27648\,\lambda v ( -2\,u+a ) {\mu}^{3}+27648\,\lambda v (
            b+{a}^{2} ) ( u+a ) ) U V+ ( 49152\,{\lambda }^{3}z (
            36\,{a}^{3}u-18\,{a}^{2}{z}^{2}+12\,{a}^{4}+9\,{z}^{2}b+30\,{b}^
            {2}\\&&%
            -12\,{a}^{2}b-18\,aub ) {\mu }^{3}+49152\,{\lambda }^{3}z (
            -6\,b+18\,ua+12\,{a}^{2}-9\,{z}^{ 2} ) {\mu }^{6}+49152\,{\lambda
            }^{3}z ( b+{a}^{2} ) ^{2} ( 4\,{a}^{2}+18\,ua\\&&%
            -14\,b-9 \,{z}^{2} ) +196608\,{\lambda }^{3}{\mu }^{9}z ) U+ (
            -73728\,v{\lambda }^{2} ( b+{a}^{2} ) ^{2} ( u+a )
            -73728\,v{\lambda }^{2} ( 4\,u-8\,a ) {\mu }^{6}-73728\,v{\lambda
            }^{2 }\\&&%
            ( -4\,bu+9\,{z}^{2}a-7\,{a}^{3}-13\,u{a}^{2}+2\,ab ) {\mu }^{3} )
            V-7340032\,{\lambda }^ {4}{\mu }^{12}z-262144\,{\lambda }^{4}z (
            60\,ua-56\,b+85\,{a}^{2}-30\,{z}^{2} ) {\mu }^{9}\\&&%
            -262144 \,{\lambda }^{4}z ( b+{a}^{2} ) (
            31\,{a}^{4}+72\,{a}^{3}u-10\,{a}^{2}b-36\,{a}^{2}{z}^
            {2}+18\,aub+13\,{b}^{2}-9\,{z}^{2}b ) {\mu }^{3}-262144\,{\lambda
            }^{4}z ( b+{a}^{2} ) ^{3 }\\&&%
            ( {a}^{2}+6\,ua-5\,b-3\,{z}^{2} ) -262144\,{\lambda }^{4}z (
            15\,{b}^{2}+87\,{a}^{4}-
            63\,{a}^{2}{z}^{2}+45\,{z}^{2}b-90\,aub-33\,{a}^{2}b+126\,{a}^{3}u
            ) {\mu }^{6})/W^2;\end{array}$\;%
          $S~:= { ( -u+a ) {Y}^{2}\lambda -4\,vY-4\,a-4\,u}$;\ %
            $\Delta~:= \sqrt [3]{({Z}-S)/( \lambda {Y}^{2}-4 )}$;\;%
          $x_t~:={ {\Delta}-\mu/{\Delta}}$; $y_t~:= \lambda\,
          ({X}^{3}/2+3\,\mu X/2+a-u)\, Y-2\,v$;\;%
          \KwRet{$(x_t:y_t:1)$}
        \caption{Genus2Type2Encode}
        \label{algo:genus2-sy-rational}
  \end{algorithm2e}
\end{footnotesize}
}}
\label{fig:algo:genus2-sy-rational}
\caption{Encoding on genus 2 curves (of the type 2)}  
\end{figure}
\medskip

So, we obtain the following theorem.
\begin{theorem}\label{sec:sy-be-rational-1}
  Let $\F_q$ be the finite field with $q$ elements. Suppose $q$ odd and $q
  \equiv 2 \mod 3$. 
  Let $H_{2,\lambda,\mu,a, v,w}/\F_q~$ be the hyperelliptic curve of genus 2
  defined by Eq.~\eqref{eq:9}.

  Then, Algorithm~\ref{algo:genus2-sy-rational} computes a
  deterministic encoding $e_{2, \lambda,\mu,a, v,w}: \F_q^*\setminus
  {\mathcal S}_2 \rightarrow H_{2,\lambda,\mu,a, v,w}$, where ${\mathcal S}_2$ is a
  subset of $\F_q$ of size at most $233$, in time $\OO(\log^{2+o(1)}
  q)$.
\end{theorem}

\begin{myproof}
The previous formulae defines a deterministic encoding provided that
$t$, $W$, ${\lambda Y^2 - 4}$ and $Z-S$ are not 0. 

The condition $W=0$ yields a polynomial in $U$ of degree $2$, we thus
have at most 2 values for $U$ for which $W=0$. Each value of $U$ then
yields a polynomial in $t$, derived from $\delta$, of degree $8$. We
thus have at most $16$ values for $t$ to avoid in this case.

The condition ${\lambda Y^2 - 4}=0$ similarly yields 2 values for $Y$.
Each such value yields in return a polynomial of degree $2$ in $U$,
and degree 1 in $V$, which can be seen as a curve in $t$ and
$\tau=\sqrt [3]{2\,t\,\delta}$ of degree at most $6$. Besides $\tau^3
= 2\,t\,\delta$ is a curve of degree at most $9$. Bezout's theorem
yields thus a maximal number of $2\times 6\times 9 = 108$ intersection
points, or equivalently values for $t$, to avoid in this case.

Finally, the condition $Z=S$ can be seen as a curve in $t$ and $\tau$
of degree $12$. Thus, this yields a maximal number of $12 \times 9 =
108$ values too.

So, the total number of field elements which cannot be encoded finally
amounts to at most $1+16+2\times 108 = 233$.
\end{myproof}

\subsubsection{Cardinality of the image.}

Let $(X, Y)$ be a rational point on $H_{2, \lambda,\mu,a, v,w}$ and
$t$ a preimage by $e_{2, \lambda,\mu,a, v,w}$. Then we have seen in
the proof of Theorem~\ref{sec:sy-be-rational-1} that $t$ and
$\tau=\sqrt [3]{2\,t\,\delta}$ are defined as intersection points of
two curves, one of degree $6$ parameterized by $Y$ and the other one
of degree $9$ from the definition of $\delta$. In full generality,
this might yield for some curves and some of their points a total
number of at most $54$ $t$'s.  Therefore, $|\IM e_{1, a, b}| \geqslant
({q-233})/{54}$.

\subsubsection{Number of curves.}
Igusa invariants of these curves are equal to
\begin{small}
  \begin{displaymath}
    \begin{array}{rcl}
      J_2 &=& -2^6\,3\,{\lambda }^{2} ( 9\,{\mu }^{3}+9\,{a}^{2}+10\,b )\,,\\%
      J_4 &=& 2^9\,3\,b{\lambda}^{4} ( 297\,{\mu }^{3}+54\,{a}^{2}+55\,b )\,,\\%
      J_6 &=& 2^{14}\,{b}^{2}{\lambda}^{6} ( -6480\,{\mu }^{3}+81\,{a}^{2}+80\,b)\,,\\%
      J_8 &=& -2^{16}\,3\,{b}^{2}{\lambda}^{8} ( 31347\,{\mu
      }^{6}-134136\,{\mu }^{3}{a}^{2}-158310\,b{\mu }^{3}+ 
      11664\,{a}^{4}+23940\,b{a}^{2}+12275\,{b}^{2} )\,,\\%
      J_{10} &=& -2^{24}\,3^6\,{b}^{3}{\lambda}^{10} ( {\mu }^{6}+2\,{\mu
      }^{3}{a}^{2}-2\,b{\mu }^{3}+{a}^{4}+2 
      \,b{a}^{2}+{b}^{2} )\,.
\end{array}
  \end{displaymath}
\end{small}
Here, the geometric locus of these invariants is a surface of dimension
2 given by a homogeneous equation of degree 30,
\begin{footnotesize}
  \begin{multline*}
    11852352\,{J_{2}}^{5}{J_{10}}^{2}+196992\,{J_{2}}^{5}J_{4}\,J_{6}\,J_{10}-362998800\,{J_{2}}^{3}J_{4}\,{J_{10}}^{2}+64\,{J_{2}}^{6}{J_{6}}^{3}-636672\,{J_{2}}^{4}{J_{6}}^{2}J_{10}\\
-895349625\,{J_{2}}^{2}J_{6}\,{J_{10}}^{2}-64097340625\,{J_{10}}^{3}-373248\,{J_{2}}^{4}{J_{4}}^{3}J_{10}-4466016\,{J_{2}}^{3}{J_{4}}^{2}J_{6}\,J_{10}\\
+2903657625\,J_{2}\,{J_{4}}^{2}{J_{10}}^{2}-3984\,{J_{2}}^{4}J_{4}\,{J_{6}}^{3}+606810\,{J_{2}}^{2}J_{4}\,{J_{6}}^{2}J_{10}+3383973750\,J_{4}\,J_{6}\,{J_{10}}^{2}+1647\,{J_{2}}^{3}{J_{6}}^{4}\\+49583475\,J_{2}\,{J_{6}}^{3}J_{10}+11290752\,{J_{2}}^{2}{J_{4}}^{4}J_{10}+38072430\,J_{2}\,{J_{4}}^{3}J_{6}\,J_{10}+76593\,{J_{2}}^{2}{J_{4}}^{2}{J_{6}}^{3}\\-115457700\,{J_{4}}^{2}{J_{6}}^{2}J_{10}+20196\,J_{2}\,J_{4}\,{J_{6}}^{4}-530604\,{J_{6}}^{5}-85386312\,{J_{4}}^{5}J_{10}-468512\,{J_{4}}^{3}{J_{6}}^{3}\,.
\end{multline*}
\end{footnotesize}
This shows that Eq.~\eqref{eq:9} defines $\OO(q^2)$ distinct curves over $\F_q$.

\section{Hyperelliptic curves of
  any genus}
\label{sec:determ-param-hyper}

In this section, we present two families of parametric polynomials
which provide deterministic parameterizable hyperelliptic curves of
genus $g\geqslant 2$. 

\subsection{Quasiquadratic polynomials}
\label{sec:quas-polyn}

Curves of the form $y^2= f(x^d)$ where $f$ is a family of solvable polynomials
whatever is its constant coefficient may yield parameterizable hyperelliptic
curves.  Typically, we may consider polynomials $f$ of degree 2, 3 or 4 or
some solvable families of higher degree polynomials. Here, we restrict ourselves to
quadratic polynomials since it yields non trivial hyperelliptic curves for any
genus.\medskip

We define quasiquadratic polynomials as follows.
\begin{definition}[Quasiquadratic polynomials]
  Let $\K$ be a field and $d$ be an integer coprime with $\mathrm{char
  }~\K$.  The family of quasiquadratic polynomials $q_{a,b}(x) \in
  \K[x]$ of degree $2d$ is defined for $a,b \in \K$ by
  \begin{math}
    q_{a,b}(x)  = x^{2d} + a x^{d} + b\,.
  \end{math}
\end{definition}

Quasiquadratic polynomials define an easily parameterized family of
hyperelliptic curves $y^2=q_{a,b}(x)$ (see
Algorithm~\ref{algoquasiquad}). When $d$ does not divide $q-1$, these curves
are isomorphic to curves $y^2=q_{1,a}(x)$ by the variable substitution
$x\rightarrow a ^{1/d}x$.

\begin{figure}[htbp]
  \centering
  {\parbox{0.6\textwidth}{%
      \begin{footnotesize}%
        \begin{algorithm2e}[H]%
          \SetKwInOut{Input}{input} \SetKwInOut{Output}{output} %
          \Input{A curve $H_{a}: x^{2d} + x^d + a = y^2$, and $t \in
            \F_q\setminus\{1/2\}$.}%
          \Output{A point $(x_t:y_t:1)$ on $H_{a}$}%
          \BlankLine%
          $\alpha~:= ({{t}^{2}-a})/({1-2\,t})$\;%
          $x_t~:= \alpha^{1/d}$;\ $y_t~:= ({-a+t-{t}^{2}})/({1-2\,t})$\;%
          \KwRet{$(x_t:y_t:1)$}%
          \label{algoquasiquad}
          \caption{QuasiQuadraticEncode}
        \end{algorithm2e}
      \end{footnotesize}
    }}
  \caption{Encoding on quasiquadratic curves}
  \label{fig:algoquasiquad}
\end{figure}

\begin{theorem}
  Let $\F_q$ be the finite field with $q$ elements. Suppose $q \not =
  2,3$ and $d$ coprime with $q-1$.
  Let $H_{a}/\F_q: y^2= x^{2d} + x^{d} + a$ be an hyperelliptic curve where
  $a$ is such that the quasiquadratic polynomial $q_{1,a}$ has a non-zero
  discriminant over $\F_q$.

  Algorithm~\ref{algoquasiquad} computes a deterministic encoding
  $e_{a}: \F_q^*\setminus \{1/2\} \rightarrow H_{a}$ in time
  $\OO(\log^{2+o(1)} q)$.

\end{theorem}

\paragraph{Genus of $H_{a}$.}
\label{sec:genus-dimension-quasiquadr}
Let $q_{1,a} \in \F_q[X]$ and $H_{a}:q_{1,a}(x) = y^2$, where $q_{1,a}$ has
degree $2d$. We have requested that the discriminant of $q_{1,a}$ is not
$0$. This implies that $q_{1,a}$ has exactly $2d$ distinct roots. Thus $H_{a}$
has genus $d-1$ provided $H_{a}$ has no singularity except at the point at
infinity.

It remains to study the points of the curve where both derivatives in $x$ and
$y$ are simultaneously $0$. This implies $y = 0$. Thus the only singular
points are the common roots of $q_{1,a}(x)$ and its derivative. Since we
request that the discriminant of $q_{1,a}$ is not $0$, there are no singular
point.\medskip

For $d=3$, $H_{a}$ is the well known family of genus 2 curves with
automorphism group $D_{12}$~\cite{cardonaquer07}. The geometric locus of these
curves is a one-dimensional variety in the moduli space. Moreover, when $x
\rightarrow x^d$ is invertible over $\F_q$, these curves all have exactly $q+1$
$\F_q$-points (but they have a much better distributed number of
$\F_{q^2}$-points).

\paragraph{The encoding.}
\label{sec:encoding-1} 
The parameterization is quite simple.  Let $H_{a}: x^{2d} + x^d + a = y^2$ be a
quasiquadratic hyperelliptic curve. Setting $x = \alpha^{1/d}$ reduces the
parameterization of $H_{a}$ to the parameterization of the conic $\alpha^2 +
\alpha + a-y^2 = 0$, which easily gives $\alpha = (-a+{t}^{2})/({1-2\,t})$ and
$y = ({-a+t-{t}^{2}})/({1-2\,t})$ for some parameter $t$. We finally obtain
Algorithm~\ref{algoquasiquad}.

\paragraph{Cardinality of the image.}
\begin{theorem}
  Given a rational point $(x:y:1)$ on $H_{a}:q_{1,a}(x) = y^2$, the equation
  $e_{a}(t) = (x:y:1)$ has exactly 1 solution. Thus,
  $|\IM e_{a}| = q-1$
\end{theorem}

\begin{myproof}
  Let $\alpha = x^d$, then $t$ is a solution of the degree 1 equation
  $y + \alpha = {ta}/({a-2t})$. 
\end{myproof}

\subsection{De Moivre's polynomials}
\label{sec:demoivres-family}
This well-known family of degree 5 polynomials was first introduced by
De Moivre for the study of trigonometric equalities and its study in a
Galoisian point of view was done by Borger in \cite{1908:Borger}. This
definition can be easily generalized for any odd degree.

\begin{definition}[De Moivre's polynomials]
  Let $\K$ be a field and $d$ be an odd integer coprime with
  $\mathrm{char }~\K$.  The family of De Moivre's polynomials $p_{a,b}(x)
  \in \K[x]$ of degree $d$ is defined for $a,b \in \K$  by
\begin{displaymath}
  p_{a,b}(x)  = x^d + d a x^{d-2} + 2da^2x^{d-4} + 3da^3x^{d-6} +\cdots + 2 d a^{(d-1)/2-1} x^3 +
  da^{(d-1)/2}x + b\,.
\end{displaymath}
\end{definition}

\noindent
\textit{Examples.}
De Moivre's polynomials of degree $5$ are $x^5 + 5ax^3 +
5a^2x + b$.
De Moivre's polynomials of degree $13$ are $x^{13} + 13 a
x^{11} + 26 a^2 x^9 + 39 a^3 x^7 + 39 a^4 x^5 + 26 a^5 x^3 + 13 a^6 x
+ b$.
\medskip

Borger proved in \cite{1908:Borger} that De Moivre's polynomials of
degree $5$ are solvable by radical, the same is true for De Moivre's
polynomials of any degree.
\begin{lemma}[Resolution of De Moivre's polynomials]
  Let $p_{a,b}$ be a De Moivre's polynomial of degree $d$, let $\theta_0$ and
  $\theta_1$ be the roots of $q_{a,b}(\theta) = \theta^2 + b\theta - a^d$,
  then the roots of $p_{a,b}$ are 
  \begin{displaymath}
    (\omega_k \theta_0^{1/d} + \omega_k^{d-1}
    \theta_1^{1/d})_{0 \leqslant k < d}  
  \end{displaymath}
  where $(\omega_k)_{0 \leqslant k < d}$ are the $d$-th roots of unity.
  \label{sec:demo-polyn}
\end{lemma}
\begin{myproof}
  As in the case of degree 5 (see \cite{1908:Borger}), we do the
  variable substitution $x = \gamma - a/\gamma$, then $\gamma^d$ is a
  root of the polynomial $q_{a,b}(\theta)$.
\end{myproof}

De Moivre's polynomials also define a family of deterministically
parameterized hyperelliptic curves for any genus. 

\begin{figure}[htbp]
  \centering
  {\parbox{0.85\textwidth}{%
    \begin{footnotesize}%
      \begin{algorithm2e}[H]
        \SetKwInOut{Input}{input} \SetKwInOut{Output}{output} %
        \Input{A curve $H:p_{a,b}(x) - y^2 = 0$, $u_0, v_0 \in \F_q$ such that
          $4 a^5 + b^2 -2bu_0 + u_0^2 = v_0^2$ and $t \in
          \F_q^*\setminus{{\mathcal S}}$.}%
        \Output{A point $(x_t:y_t:1)$ on $H$}%
        \BlankLine%
        $\delta~:= -({3a^d+b^2+t^4})/{6t} - 2\,b^3/27 - a^d b/3 - t^6/27$;\ $A
       ~:= \delta^{1/3 \bmod q-1} + t^2/3$\;%
        $Y~:= tA - ({3a^d+b^2+t^4})/({6t})$\;%
        $\alpha~:= {3a^d}/({-3A+b})$\;%
        $y_t~:= {-3 Y}/({-3A + b})$;\ $x_t~:= \alpha^{1/d \bmod q-1} +
        ({-a^d}/{\alpha})^{1/d \bmod q-1}$\;%
        \KwRet{$(x_t: y_t:1)$}%
    \label{EncodingDeMoivreHighChar}
    \caption{DeMoivreEncode}
    \end{algorithm2e}
  \end{footnotesize}
}}
\label{fig:EncodingDeMoivreHighChar}
\caption{Encoding on De Moivre's curves}
\end{figure}

\begin{theorem}
  Let $\F_q$ be the finite field with $q$ elements. Suppose $q$ odd and $q
  \equiv 2 \mod 3$ and $d$ coprime with $q-1$. 
  Let $H_{a,b}/\F_q: y^2=p_{a,b}(x)$ be the hyperelliptic curve where
  $p_{a,b}$ is a De Moivre polynomial defined over $\F_q$ with non-zero
  discriminant.

  Algorithm~\ref{EncodingDeMoivreHighChar} computes a deterministic encoding
  $e_{a,b}: \F_q^*\setminus {\mathcal S} \rightarrow H_{a,b}$, where
  ${\mathcal S}$ is a subset of $\F_q$ of size at most $7$, in time
  $\OO(\log^{2+o(1)} q)$.

\end{theorem}

Conversely, given a point on $H$ we study how many elements in
$\F_q$ yield this point.

\begin{theorem}
  Given a point $(x:y:1) \in H_{a,b}(\F_q)$, we can compute the solutions $s$ of the
  equation $e_{a,b}(s) = (x:y:1)$ in time $\OO(\log^{2+o(1)} q)$. There are at
  most $8$ solutions to this equation. 
\end{theorem}

We give below proofs of these two theorems.

\ifforarxiv
\subsubsection{Finite fields of odd characteristic.}
\label{sec:finite-fields-odd}
\fi

\paragraph{Genus and dimension of $H_{a,b}$.}
\label{sec:genus-dimension-hab}
As in Section~\ref{sec:genus-dimension-quasiquadr}, since we request the
discriminant of $q_{a,b}$ to be nonzero, there is no singularity except the
point at infinity. Thus the genus of $H_{a,b}$ is $(d-1)/2$.

\paragraph{The encoding.}
\label{sec:encoding}
Thanks to Lemma~\ref{sec:demo-polyn}, parameterizing rational points
on $H_{a,b}: p_{a, b}(x) = y^2$ amounts to finding roots of $\theta^2
+ (b-y^2) \theta - a^d$. Let them be $\alpha, \alpha'$, then we have
$x = \alpha^{1/d} + \alpha'^{1/d}$, $\alpha \alpha' = -a^d$ and
$\alpha + \alpha' = y^2-b$. Thus $\alpha^2 - a^d = \alpha y^2
-b\alpha$. This is a genus 1 curve with variable $\alpha, y$ which is
birationally equivalent to $Y^2 = A^3 + (-a^d -\frac{1}{3}b^2)A +
\frac{2}{27} b^3 + \frac{1}{3}a^d b$, with $\alpha = {3a^d}/({-3A+b})$
and $y = {-3 Y}/({-3A + b})$.

This curve can be parameterized with Icart's method. This yields $ A =
\sqrt[3]{\delta} + t^2/3,\ \ Y = tA - ({3a^d+b^2+t^4})/{6t} $ where $ \delta =
-5{3a^d+b^2+t^4})/{6t} - {2}\, b^3/27 - a^d b/3 - t^6/27$\,.
We finally obtain Algorithm~\ref{EncodingDeMoivreHighChar}. 

\paragraph{Restrictions.}
Previous necessary conditions on an encoding are also sufficient to
give an encoding for $t \in \F_q$ provided that every variable
substitution is computable. 

In order to compute $A$ and $Y$ using the encoding from
\cite{DBLP:conf/crypto/Icart09}, we need $t \not = 0$. Then computing
$y$ and $\alpha$ from $A$ and $Y$ we also request $-3A + b \not = 0$,
that is $\delta \not = ({b}/{3} - {t^2}/{3})^3$. This amounts
to a degree 7 equation, thus at most 7 elements of $\F_q$ are not
encodable. 

\paragraph{Complexity.}
Our encoding function uses one Icart's encoding, of complexity
$\OO(\log^{2+o(1)} q)$ operations in $\F_q$, two exponentiations for computing
$d$-th roots and a constant number of field operations. The total
amounts to $\OO(\log^{2+o(1)} q)$ running time.

\paragraph{Computation of $e_{a,b}^{-1}$.}
Let $(x:y:1)$ be a point on $H_{a,b}$. The polynomial $\beta^2 + x\beta
- \sqrt[5]{(-a^d)}$ has at most two roots. Let $\beta$ be one, and
$\alpha = \beta^5$. Let then $A = 1-3(b\alpha - 3a^d)/\alpha$ and $Y =
-y a^d/\alpha$, we are reduced to finding the solutions of an Icart's
encoding. It admits at most $4$ solution per $\alpha$, thus there are
at most 8 solutions to the equation $e_{a, b}(t) = (x:y:1)$.

\paragraph{Genus 2 case.}
In this case we are interested in the dimension of the family of
curves defined by De Moivre's polynomials, $H:y^2= x^5 + 5ax^3 +
5a^2x + b$.  We have computed their Igusa invariants,
\begin{multline*}
  J_2  =  700\,{a}^{2}\,,\ 
  J_4  =  13750\,{a}^{4}\,,\ 
  J_6  =  -2500\,a ( 3\,{a}^{5}+32\,{b}^{2} )\,,\\
  J_8  =  -15625\,{a}^{3} ( 3109\,{a}^{5}+896\,{b}^{2} )\,,\ 
  J_{10}  =  800000\, ( 4\,{a}^{5}+{b}^{2} ) ^{2}\,,
\end{multline*}
from which it is easy to derive numerous algebraic relations. This reduces the
set of curves from an expected $q^2$ because of the two parameters $a$ and $b$
to a set of cardinality $\OO(q)$.

\ifforarxiv
\subsubsection{Finite fields of characteristic two.}
\label{sec:finite-fields-small}

The case of characteristic 2 is very similar. De Moivre's polynomials are
solvable using the same auxiliary polynomial. A dimension $1$ family of genus
2 curves is given by $p_{a,b}(x) = y+y^2$ which are also $p_{a,b+y+y^2}(x) =
0$.

\begin{figure}[htbp]
  \centering
  {\parbox{0.95\textwidth}{%
    \begin{footnotesize}%
      \begin{algorithm2e}[H]%
        \SetKwInOut{Input}{input} \SetKwInOut{Output}{output} %
        \Input{A curve $H:p_{a,b}(x) - y - y^2 = 0$ on $\F_q$ with $q$ even
          and $t \in \F_q^*\setminus{{\mathcal S}}$.}%
        \Output{A point $(x_t:y_t:1)$ on $H$}%
        \BlankLine%
        Reduce the elliptic curve $E: \alpha^2 + y^2 \alpha + b \alpha + a^5
        = 0$ to the Weierstrass form $\alpha^2 + y \alpha = y^3 + cy+d$\;%
        Encode $t$ on $E$ and obtain the point $(\alpha_t, y_t)$\;%
        $x_t~:= \alpha_t^{1/5 \bmod q-1} + a/\alpha_t^{1/5 \bmod q-1}$\;%
        \KwRet{$(x_t: y_t:1)$.}%
        \label{EncodingDeMoivreSmallChar} \caption{DeMoivreEncodeChar2}
    \end{algorithm2e}
  \end{footnotesize}
}}

\caption{De Moivre's encoding in even characteristic}
\label{fig:EncodingDeMoivreSmallChar}
\end{figure}

\begin{theorem}
  Let $\F_q$ be the finite field with $q$ elements. Suppose $q$ even, $q
  \equiv 2 \mod 3$ and let $d$ odd coprime with $q-1$.
  Let $H_{a,b}/\F_q: y^2 + y=p_{a,b}(x)$ be an hyperelliptic curve where
  $p_{a,b}$ is a De Moivre's polynomial defined over $\F_q$ with non-zero
  discriminant.

  Algorithm~\ref{EncodingDeMoivreSmallChar} computes a deterministic
  encoding $e_{a,b}: \F_q^*\setminus {\mathcal S} \rightarrow H_{a,b}$, where ${\mathcal S}$ is a
  subset of $\F_q$ of size at most $12$, running in time $\OO(\log^{2+o(1)} q)$.

\end{theorem}

\paragraph{Proof.}

Recall that $H: p{a,b - y - y^2} = 0$. We consider the
auxiliary equation $\theta^2 + (b - y - y^2) \theta + a^d = 0$. Let
$\alpha_0$ be a root of this equation, then the second root is
$\alpha_1 = a^d / \alpha_0$. Suppose $\alpha_0$ parameterized, then
the (unique) root of our $p_{a, b - y - y^2}$ De Moivre's polynomial is
$x = \sqrt[5]{\alpha_0} + \sqrt[5]{\alpha_1}$. We are reduced to the
problem of parameterizing $y$ and $\alpha_0$.

Remark that $b - y - y^2 = \alpha_0 + \alpha_1$. This implies that $y$ and
$\alpha_0$ lie on the genus 1 curve $E: \alpha_0^2 + y^2 \alpha_0 + b
\alpha_0 + a^5 = 0$. This curve can be easily parameterized
using~\cite{DBLP:conf/crypto/Icart09}. 

\fi

\subsection{Encoding into the Jacobian of an hyperelliptic curve}

Let $H$ be a genus $g$ hyperelliptic curve defined over a finite field $\F_q$
coming from the families defined in the previous sections
\ref{sec:genus-2-case}, \ref{sec:quas-polyn} and \ref{sec:demoivres-family}. 
We provide deterministic functions $e_H$ which construct rational points on
$H$ from elements in $\F_q\setminus {\mathcal S}$, where ${\mathcal S}$ is a small subset of $\F_q$
which depends on the definition of $H$.  In this section, we present two
straightforward strategies for encoding divisors in $\J_{H}(\F_q)$ the
Jacobian of $H$.

Recall that each class in $\J_{H}(\F_q)$ can be uniquely
represented by a reduced divisor.
A divisor $D$ is said to be reduced when it is a formal sum of points
$\sum_{i=1}^r P_i - rP_{\infty}$ with $r\leqslant g$, $P_i\not=-P_j$ for
$i\not= j$ and this sum is invariant under the action of the Galois
group $\operatorname{Gal}(\overline{\F}_q/\F_q)$.

\paragraph{Encoding $1$-smooth reduced divisors.}
There is a particular subset, denoted by $\mathcal{D}_1$, of reduced divisors
which are called $1$-smooth. These divisors are the ones with only rational
points in their support. From our encoding function $e_H$, one easily deduces a
function providing elements in $\mathcal{D}_1$: in a first step, a set of
$r\leqslant g$ points (none of these points in this set is the opposite of
another one) is produced then a divisor is constructed from this set. This first
step can be done deterministically by computing $g$ points with $e_H$ and
eliminating possible collisions after negation. When $q$ is large enough, the
proportion of $\mathcal{D}_1$ in $\J_{H}(\F_q)$ is $\approx 1/g!$ moreover,
since $e_H$ is not surjective, this function may be not surjective too.  If one
wants to construct more general reduced divisors, another strategy has to be
used.

\paragraph{Extension of the base field and encoding.}
In the definition of the encoding $e_H$, we assume specific conditions
on the base field $\F_q$ so that some power functions are
deterministically bijective. If one wants to directly encode in the
Jacobian of an hyperelliptic curve $H$ defined over $\F_q$, one can
change the conditions in the following way. These specific conditions
are now assumed for the extension field $\F_{q^g}$ (and thus no more
on $\F_q$). The function $e_H$ becomes an encoding $e'_H$ from
$\F_{q^g}\setminus {\mathcal S}^{\prime}$ (where the set ${\mathcal S}^{\prime}$ can be
computed in the same manner as ${\mathcal S}$) to the set of $\F_{q^g}$-rational
points of $H$. From this new function $e'_H$ one can compute a set of
$k$ points in $H(\F_{q^g})$ such that the sum of their degree over
$\F_q$ is less than $g$. By constructing the $\F_q$-conjugates of
these points and eliminating the possible collision after negation, we
deduce a reduced divisor of $\J_{H}(\F_q)$. This second strategy is
more general than the former but it does not assume the same
conditions on the field $\F_q$.

Remark that these two encodings are clearly ``weak encoding'' in the
sense of \cite{conf/crypto/BrierCIMRT10}.

\section{Conclusion and future work}
We have almost extensively studied families of genus 1 and 2 curves
which admit a deterministic algebraic encoding using the
resolution of a degree 3 polynomial. We come to a new encoding map for
Hessian elliptic curves and we give, for the first time to our
knowledge, encoding maps for large families of genus 2 curves.  We
have also sketched families of higher genus hyperelliptic curves whose
deterministic algebraic parameterization is based on solvable
polynomials of higher degree arising from K\"ummer theory.

On-going work is being done to extend these families to finite fields of small
characteristic. A natural question is to generalize the method to solvable
degree 5 polynomials too, in the hope to first find a deterministic algebraic
parameterization of every genus 2 curve, then of families of higher genus
curves. 

\bibliographystyle{plain}

\bibliography{hinec}

\begin{thebibliography}{10}

\bibitem{Atkin:1993:ECP}
A.~O.~L. Atkin and F.~Morain.
\newblock {Elliptic curves and primality proving}.
\newblock {\em Mathematics of Computation}, 61(203):29--68, July 1993.

\bibitem{BeKoLa2010}
D.~J. Bersntein, D.~Kohel, and T.~Lange.
\newblock {Twisted Hessian curves}.
\newblock \url{http://www.hyperelliptic.org/EFD/g1p/auto-twistedhessian.html}.

\bibitem{BonFra2001}
D.~Boneh and M.~Franklin.
\newblock {Identity-Based Encryption from the {Weil Pairing}}.
\newblock In Joe Kilian, editor, {\em Advances in Cryptology -- {CRYPTO} '
  2001}, volume 2139 of {\em Lecture Notes in Computer Science}, pages
  213--229. Springer-Verlag, Berlin Germany, 2001.

\bibitem{1908:Borger}
R.~L. Borger.
\newblock On {D}e {M}oivre's quintic.
\newblock {\em The American Mathematical Monthly}, 15(10):171--174, 1908.

\bibitem{Magma:97}
W.~Bosma, J;~J. Cannon, and C.~Playoust.
\newblock The {M}agma {A}lgebra {S}ystem {I}: The user language.
\newblock {\em J. Symb. Comput.}, 24(3/4):235--265, 1997.

\bibitem{conf/crypto/BrierCIMRT10}
E.~Brier, J.-S. Coron, T.~Icart, D.~Madore, H.~Randriam, and M.~Tibouchi.
\newblock Efficient indifferentiable hashing into ordinary elliptic curves.
\newblock Cryptology ePrint Archive, Report 2009/340, 2009.
\newblock \url{http://eprint.iacr.org/2009/340/}.

\bibitem{cardonaquer07}
G.~Cardona and J.~Quer.
\newblock {Curves of genus 2 with group of automorphisms isomorphic to ${D}_8$
  or ${D}_{12}$}.
\newblock {\em Trans. Amer. Math. Soc.}, 359:2831--2849, 2007.

\bibitem{Cox:GaloisBook}
D.~A. Cox.
\newblock {\em {G}alois theory}.
\newblock Pure and Applied Mathematics (New York). Wiley-Interscience [John
  Wiley \& Sons], Hoboken, NJ, 2004.

\bibitem{FaJo2010}
R.~R. Farashahi and M.~Joye.
\newblock {Efficient Arithmetic on Hessian Curves}.
\newblock In {\em Public Key Cryptography - PKC 2010}, volume 6056 of {\em
  Lecture Notes in Computer Science}, pages 243--260. Springer, 2010.

\bibitem{DBLP:conf/crypto/Icart09}
T.~Icart.
\newblock {How to Hash into Elliptic Curves}.
\newblock In Shai Halevi, editor, {\em CRYPTO}, volume 5677 of {\em Lecture
  Notes in Computer Science}, pages 303--316. Springer, 2009.

\bibitem{maple}
Waterloo~Maple Incorporated.
\newblock Maple.
\newblock \url{http://www.maplesoft.com/}.
\newblock Waterloo, Ontario, Canada.

\bibitem{1564550}
J.~R. Sendra, F.~Winkler, and S.~Prez-Diaz.
\newblock {\em {Rational Algebraic Curves: A Computer Algebra Approach}}.
\newblock Springer Publishing Company, Incorporated, 2007.

\bibitem{DBLP:conf/ants/ShallueW06}
A.~Shallue and C.~van~de Woestijne.
\newblock {Construction of Rational Points on Elliptic Curves over Finite
  Fields}.
\newblock In Florian Hess, Sebastian Pauli, and Michael~E. Pohst, editors, {\em
  ANTS}, volume 4076 of {\em Lecture Notes in Computer Science}, pages
  510--524. Springer, 2006.

\bibitem{ulas07}
M.~Ulas.
\newblock {Rational points on certain hyperelliptic curves over finite fields}.
\newblock {\em Bull. Polish Acad. Sci. Math.}, (55):97--104, 2007.

\end{thebibliography}

\end{document}